\documentclass[5p,preprint]{elsarticle}

\usepackage{hyperref}
\usepackage{amsfonts,amsmath,amssymb} 
\usepackage{graphicx,graphics,color}

\biboptions{sort&compress}

\journal{Physics of the Dark Universe}

\usepackage{amssymb}

\ead{email address}
\def\mnras{Mon. Not. Roy. Astron. Soc.} 

\def\mnras{MNRAS} \def\apj{ApJ} \def\apjl{ApJ}

  \def\prd{Phys. Rev. D}
 \def\prl{Phys. Rev. Lett.} 
 \def\nat{Nature} 
\def\pasj{PASJ}  

       \def\jcap{J.
Cosm. Astropart. Phys.} \def\aj{Astron. J.}

\begin{document}

\begin{frontmatter}

\title{Constraining light fermionic dark matter with binary pulsars}

\author{L. Gabriel G\'omez}
\address{Group of Research on Relativity and Gravitation\\ Escuela de F\'isica, Universidad Industrial de Santander\\
Ciudad Universitaria, Bucaramanga 680002, Colombia}
\ead{gabrielphysics@gmail.com}





\begin{abstract}
A binary system embedded in a Dark Matter (DM) background may experience a change in its orbital period due to dynamical friction as the binary moves through a wind of DM particles. Since the orbital motion variation depends on the DM environment, such a phenomenon provides an intriguing way of constraining the properties of DM from timing pulsar observations. We compute such a perturbative effect on the binary evolution considering that DM is constituted of degenerate gas of free fermions. The analysis point out that the secular change of the orbital period is more sensitive, and likely measurable, to degenerate fermions with masses $\gtrsim50$~ eV, depending slightly, but still being distinguishable, on the binary star configuration (e.g.  NS-NS,  NS-WD  and  WD-WD). Interestingly, we find that NS-NS binary systems with large orbital periods, $P_{b}\gtrsim100$~days, experience larger orbital period decays. We also show that this effect is clearly increased, under the former conditions, in binaries orbiting small DM halos, which correspond to extragalactic pulsars. This situation represents the best astrophysical scenario to test such effects of light fermionic DM. We use some available measurements of the orbital period time-derivative for long-period binaries in the Milky-Way to quantify more realistically this effect. For instance, measurements of the J1713+0747 pulsar set an upper bound on the fermion mass of $m_{f}\lesssim 1$~keV. This bound can be considerably improved by using pulsar timing observations of extragalactic pulsars. Under this perspective, high precision of timing pulsar observations will reveal whether DM dynamical friction effect may be tested with the upcoming generation of surveys leading to the possibility of constraining more strongly the properties of light fermionic DM. 
\end{abstract}

\begin{keyword}
Dark matter: degenerate gas of free fermions. Binary pulsars: Orbital period decay
\end{keyword}

\end{frontmatter}


\section{Introduction}
Despite great efforts in both direct \cite{2017PhRvL.118b1303A,2017arXiv170806917P,2018PhRvL.121k1302A}, and indirect \cite{Aguilar:2016kjl,2017ApJ...834..110A} detection searches and colliders \cite{2017PhLB..765...11A,2017arXiv170305236C}, the nature of DM still remains unrevealed, thus demanding more sophisticated and sensitive experiments. A less conservative position on the DM problem is to assume that the  observed  non-baryonic  gravitational effects might rather be a manifestation of some sort of modified theory. See e.g Ref.~\cite{2004PhRvD..70h3509B} among the most attractive alternatives that may account for observations.

Nevertheless, the observational evidences coming from cosmological and astrophysical scales, such as Ref.~\cite{2006ApJ...648L.109C,2004ApJ...606..702T,2018arXiv180706209P} and Ref.~\cite{1970ApJ...159..379R,2013PASJ...65..118S} respectively, support convincingly the idea of a non-baryonic component in the Universe. Accordingly, there have been proposed some DM models with promising theoretical predictions; being among the most popular candidates, apart from the \emph{weakly interactive massive particles} (WIMPs)\cite{1985NuPhB.253..375S}: ultra-light bosons (see e.g. \cite{2017PhRvD..95d3541H}) and fermionic DM (e.g. sterile neutrinos \cite{1994PhRvL..72...17D}). These latter candidates have been postulated to explain, with great success, the main problems of the CDM paradigm \cite{2017Galax...5...17D}, such as the \emph{core-cusp} \cite{1994Natur.370..629M,2010AdAst2010E...5D} and the \emph{too big to fail} \cite{2011MNRAS.415L..40B,2014MNRAS.444..222G} problems.

Non-interacting Fermionic DM models have been advocated for the formation of cores\footnote{Such a core corresponds to a polytropic core with index $n=3/2$. See for instance Refs.~\cite{2015PhRvD..91f3531C,2016arXiv160607040A}.} at the center of galaxies due to  the degeneracy pressure which prevents gravitational collapse. This unequivocal consequence of the Pauli exclusion principle has been explored, with great astrophysical interest, in order to describe adequately the observed cores in dwarf galaxies of the Milky-Way \cite{2015PhRvD..91f3531C,2016arXiv160607040A,2013APh....46...14D,2014MNRAS.442.2717D,2015MNRAS.451..622R,2015JCAP...01..002D,2017MNRAS.467.1515R}. 



As discussed, fermionic DM may play an important role in the description of cored DM halo \footnote{It is important to state that there is still a debate in the literature about the inner structure of galaxies and why some low-mass spiral galaxies are well described by cored profiles while others not \cite{2011AJ....142...24O}. Trying to reconcile such a discrepancy is a challenge for DM-only models which require inevitability the help of baryonic physics or self-interactions. For a complete discussion of the small scale crisis and unified solutions see Ref.~\cite{Tulin:2017ara,DelPopolo:2016emo}.}. We focus then on self-gravitating degenerate gas of free fermions, also known as light (sub-keV) fermions, as a simple but well motivated realization of non-interacting fermionic DM to describe the DM distribution in galaxies. For this end, we mention briefly the main advances in the development of this topic. A comprehensive attempt to describe the kinematical data of dwarf galaxies of the Milky-Way under this approach was performed in Ref.~\cite{2015JCAP...01..002D}. In there, analysis of the best-fit of the observed velocity dispersion set a narrow range for the fermion mass of $100-200$~eV,  describing thereby the structural properties of the classical satellite galaxies. A step forward was done by Ref.~\cite{2017MNRAS.467.1515R} where a more robust analysis of the kinematical data was carried out\footnote{See also Ref.~\cite{2018MNRAS.475.5385D} for further considerations in fitting the kinematical data of dwarf spheroidal galaxies with degenerate fermions.}, along with the inclusion of a thermal envelope, yielding a fermion mass lying in the range of\footnote{For a more complete discussion of how such ranges have been obtained, we refer the interested reader to the  introduction of Ref.~\cite{2017MNRAS.467.1515R}.} $70-400$~eV. We conclude this review by mentioning that analysis of rotation curves of the Milky-Way has also provided constraints on the allowed range of the fermion mass of $75-104$~eV \cite{2018arXiv181111125B}.

Despite the compelling indirect evidences for DM through its gravitational effect, new astrophysical scenarios have brought the attention as a feasible alternative to study the properties of DM. Binary pulsars have been used, for instance, to infer some information about the properties of the gravitational potential and the interstellar matter content of the Galaxy \cite{2008LRR....11....8L}. Interestingly, these objects may also have a measurable change in the orbital period, as the two stars spiral inward toward each other, which can be attributed to energy loss due to the emission of gravity waves \cite{1975ApJ...195L..51H}. These are some of the few objects with extraordinary accuracy in measurements which make them excellent tools for studying gravity (in the strong field regime) and matter around them \cite{2008LRR....11....8L}.
Furthermore, such compact-stars binaries offer an invaluable opportunity to test deviations of General Relativity though with no reported observational signals yet \cite{2006Sci...314...97K,2013Sci...340..448A}. 

Apart from this, binary pulsars may also offer interesting possibilities to know about the medium they are embedded like the intergalactic medium. For instance, dynamical friction, a drag force that is presented in the evolution of many astrophysical systems\footnote{See e.g. \cite{2008gady.book.....B} for the role of dynamical friction in the inspiral of dwarf galaxies within DM halos and the orbital evolution of black hole binaries in a stellar medium.}, may perturb the orbital motion of binary pulsar as they move through a wind of collisionless field of DM. This drag induces a \textit{wake} of medium particles on the object. As a result of this perturbative effect, due to the interaction between each binary component with its respective wake, there may be a secular change of the orbital period associated to this phenomenon \cite{1990ApJ...359..427B,2015PhRvD..92l3530P}. 

This interesting astrophysical scenario has been gaining attention as an alternative way to constrain the properties of DM from a purely gravitational interaction. The dynamical friction effect in binary pulsars has been postulated and studied in binary systems with a DM background modeled by the NFW profile \cite{2015PhRvD..92l3530P}. Recently, it has been found the conditions under which such a perturbative effect in compact-star binaries, for different DM models, may become larger than the one due to gravitational wave emission \cite{2017PhRvD..96f3001G}. Furthermore, the DM dynamical friction effects of binary pulsars has been used as probes of a Galactic DM disc scenario. \cite{2018PDU....19....1C}. Importantly, these results exhibit different theoretical predictions of orbital period decays that make DM dynamical friction a promising tool from future observations of timing pulsars to constrain or rule out DM models. 
   
As mentioned before, degenerate gas of free fermions have recently garnered attention into the astrophysical context because they yield the possibility of describing successfully the structural properties of dwarf galaxies of the Milky-Way \cite{2015JCAP...01..002D}. In view of this, the astrophysical implications of degenerate fermionic DM in galactic halos deserve further examinations. We study a novel astrophysical application of such a DM model by assessing the dynamical friction effect it produces on the orbital evolution of binary pulsars. We analyze the effect of varying  orbital parameters and the fermion mass on the secular change of the orbital period such that we can devise the best astrophysical scenario to constrain the properties of light fermionic DM. The secular change in the orbital period due to DM dynamical friction of some binary stars with measured orbital period time-derivatives is also evaluated. Thus, we show that it is possible to constrain the fermion mass by using available timing pulsar observations. For the sake of example, we use the J1713+0747 binary to set the upper limit $m_{f}\lesssim 1$~keV. Finally, we comment some prospects of timing pulsar observations of extragalactic pulsars to improve such a bound in the near future that makes our results testable predictions.

We outline now the general structure of this paper. Secs.~(\ref{sec:2})-(\ref{sec:4}) are devoted to review the underlying approach of the presented astrophysical application. We start by describing in Sec.~(\ref{sec:2}) the generalities of the fully degenerate fermions in DM halos. We proceed in Sec.~(\ref{sec:3}) with the  definition of the Chandrasekhar's dynamical friction formula. With this at hand, we present in Sec.~(\ref{sec:4}) a general illustration of how dynamical friction may originate a secular change of the orbital period in binary stars. Later in the same section we consider such a degenerate gas of fermions to account for the DM background. Finally, we compute and show in Sec.~(\ref{sec:5}) changes in the orbital period in binary pulsars under the aforementioned setup. The main conclusions of this study and the observational perspectives of this work are discussed at the end of this manuscript.

\section{General description of degenerate Fermionic DM}\label{sec:2}

We are interested in the strong degeneracy limit of non-interacting fermions, i.e, at zero temperature ($T\to0$), however, we start by describing a system of self-gravitating fermions, keeping finite temperature $T$ and chemical potential $\mu$, with the Fermi-Dirac distribution function in the non-relativistic regime
\begin{equation}
    f(p)=\frac{1}{e^{1+\beta(E-\mu)}},
\end{equation}
where $\beta$ is the inverse temperature defined as $\beta=1/K_{B}T$ and $E=p^2/2m_{f}$ is the single particle kinetic energy. $K_{B}$ is the Boltzmann constant and $m_{f}$ is the fermion mass.

Now, if we assume that DM is composed of fully degenerate free fermions, then they will occupy all the quantum states with momentum lower that the Fermi pressure $p_{f}$. The Fermi pressure corresponds to the the highest  level occupied at $T=0$. In the non-relativistic regime, it is possible to write $p=m v$. Likewise, one may write the occupation number in terms of the momentum as $n=g\frac{4\pi}{h^3}\int_{0}^{p_{f}}p^{2} dp$, where $g$ denotes the particle spin degeneracy and $h$ is the Planck constant. Taking $g=2$ for our case, then $n=8\pi p_{f}^{3}/3h^{3}$ and, along with the mass density $\rho=m n$, we get the the Fermi momentum $p_{f}=(3h^{3}\rho/8\pi m_{f})^{1/3}$. With this in mind, the pressure can be computed from the definition
\begin{equation}
 p=\frac{8\pi}{3h^{3}}\int_{0}^{p_{f}}\frac{p^{4}}{\sqrt{p^{2}+m_{f}^{2}}}dp=\frac{h^{2}}{5m_{f}^{8/3}}\left(\frac{3}{8\pi}\right)^{2/3}\rho^{5/3}.   
\end{equation}
Hence, the equation of state of a degenerate Fermi gas in the non-relativistic regime follows the polytropic equation  $P=K\rho^\gamma$, with $K=\left(\frac{3}{8\pi}\right)^{2/3}\frac{h^{2}}{5 m_{f}^{8/3}}$ and polytropic index $\gamma=5/3$. 

The density profile has to satisfy both the hydrostatic equation
\begin{equation}
\frac{dP}{dr}=-\frac{GM(r)}{r^{2}}\rho(r),
\end{equation}
and the continuity equation
\begin{equation}
M(r)=\int 4\pi r^{2}\rho(r)dr,
\end{equation}
to describe a static self-gravitating system in  thermodynamic equilibrium. This set of equations can be written in the form of the Lane-Emden equation with index $n=3/2$
\begin{equation}
    \frac{1}{\xi^{2}}\frac{d}{\xi}\left(\xi^{2}\frac{d\theta}{d\xi}\right)=-\theta(\xi)^{3/2}.
\end{equation}
We have followed closely the scaled numerical solution of this equation given by Ref.~\cite{2015JCAP...01..002D}. They introduced the dimensionless parameter\footnote{Only in this section we have used $\beta$ and $\alpha$ to denote the inverse temperature and the dimensionless parameter respectively. We use however the same letters in the remainder of this paper as angles.} $\theta$ and $\alpha$ so that\footnote{In our computations will be helpful to use, as a well approximation, the function $\rho(\xi)=\rho_{0} \cos^{3}[\frac{\pi}{8}\xi]$ found in Ref.~ \cite{2018MNRAS.475.5385D}.} $\rho(\xi)=\rho_{0}\theta(\xi)^{3/2}$ and $r=\xi\alpha$, where $\rho_{0}$ is the central density and $\alpha=(5K\rho_{0}^{-1/3}/8\pi G)^{1/2}$ with density vanishing at $R=\xi_{1}\alpha$. Making some analytical treatments, they present in particular the mass-radius relation as
 \begin{equation}
     M=-4\pi \left(\frac{5K}{8\pi G}\right)^{3}\theta^{\prime}(\xi_{1})\xi_{1}^{5} R^{-3},\label{eqn6}
 \end{equation}
where $\xi_{1}=3.65$ and $\theta^{\prime}(\xi_{1})=-0.203$ are numerical constants. It can be easily read off that the total mass scales with the fermion mass and with the radius as $M\propto m_{f}^{-8} R^{-3}$. From this latter relation is set that large self-gravitating systems can be reproduced only considering small values of $m_{f}$ and \emph{vice versa}. This model is then characterized by the two free parameters $m_{f}$ and $\rho_{0}$ we shall have in mind in all our computations.
 
On the other hand, if one assumes a gravitationally bound DM object composed of a degenerate fermion gas, there exists a densest packing of the DM phase space distribution. Hence, the phase-space density of the DM must not exceed that of the degenerate Fermi gas\footnote{This argument has been considered, since the pioneering work of Gunn \& Tremaine \cite{1979PhRvL..42..407T}, to put constraints on the fermion mass for a given DM dominated object.}. In our case, it will correspond exactly to the maximum phase space. If one demands that the Fermi velocity $v_{f}$ does not exceed the escape velocity $v_{\infty}$ of the self-gravitating degenerate gas of fermions of mass $M$ and volume $V = 4/3\pi R^{3}$ , namely $v_{f}\leq v_{\infty}$, it leads to
\begin{equation}
    \left(\frac{9Mh^3}{32\pi^2 R^{3} m_{f}^{4}}\right)^{1/3}\leq \sqrt{\frac{2GM}{R}},
\end{equation}
which sets an upper bound for the Fermi velocity that any single particle should satisfy inside the DM bound object \cite{2009JCAP...03..005BF}. We shall keep in mind this relation as a condition that any single particle, with a given velocity distribution, must satisfy inside galactic halos. 


\section{DM dynamical friction in binary stars}\label{sec:3}

We describe briefly the general scenario where an object experiences a dynamical friction as it moves through a collisionless medium of field particles. There are numerous applications of this (purely gravitational) effect  in the astrophysical context since dynamical friction is a natural force which manifests, for instance, in a change in the orbital evolution of most of the known objects (see footnote 4). This drag force induces a wake, as a result of the gravitational interaction of medium particles on the object, with a characteristic overdensity proportional to its mass \cite{2008gady.book.....B}. 

The drag force experienced by a test body of mass $m_i\gg m$, where $m$ is the DM particle mass, and with orbital velocity $v_i$ moving through the DM background characterized by the velocity distribution function $f(u)$, is given by the Chandrasekhar's dynamical friction formula \cite{1943ApJ....97..255C,2008gady.book.....B}
\begin{align} 
\textbf{f}_{fr,i}&=-4\pi G^2 m_i^2 m \left( \int_0^{\tilde{v}_i} d^3u f(u) \ln\left[\frac{b_{\rm max}}{G m_i} (\tilde{v}_i^{2}-u^2)\right] \right. \nonumber\\
& + \left. \int_{\tilde{v}_i}^{v_{\rm \infty}} d^3u f(u) \left[\ln\left(\frac{u+\tilde{v}_i}{u-\tilde{v}_i}\right)-2\frac{\tilde{v}_i}{u}\right]\right)\frac{\tilde{\textbf{v}}_i}{\tilde{v}_i^{3}},\label{a01}
\end{align} 
where the integral in the first term accounts for the fraction of particles moving slower than the object, while the integral in the second term, referring to fast particles, is limited by the escape velocity $v_{\infty}$. The above equation for the dynamical friction force considers the orbital velocity of each object with respect to the DM wind relative to the center of mass of the binary system $\tilde{\textbf{v}}_{i}=\textbf{v}_{i}+\textbf{v}_{w}$, with $\textbf{v}_{w}=v_{w}(\cos{\alpha} \sin{\beta},\sin{\alpha}\sin{\beta},\cos{\beta})$ and $\beta$ and $\alpha$ being the angles between the wind velocity vector and the perpendicular axis of the binary orbital plane and the projection of the wind velocity vector with an axes lying in the orbital plane, respectively. We follow the discussion presented in Ref.~\cite{2017PhRvD..96f3001G} in which they present at least two different contributions of wind velocities: bound and unbound binaries to the galaxy potential depending on the magnitude of the kick velocity in binaries with NS components. Thus, for the sake of generality of our results, we assume $\textbf{v}_{w}=\textbf{v}_{\rm rot}+\textbf{v}_{T}$ so that we take values for it ranging from 10~km~s$^{-1}$ all the way to 1000~km~s$^{-1}$. Here $\textbf{v}_{T}$ is the transversal velocity of the system.
%


A crucial ingredient in the Chandrasekhar's dynamical friction formula (Eqn.~(\ref{a01})) is the Coulomb logarithm which accounts for the finite size of the system and is usually defined as the ratio of the maximum and minimum impact parameters for encounters, respectively $b_{\rm max}$ and $b_{\rm min}$, i.e.
\begin{equation}
\log{\Lambda}\equiv\log{\left(\frac{b_{\rm max}}{b_{\rm min}}\right)}.\label{a03}
\end{equation}
$b_{\rm max}$ is assumed typically of the order of the size of the system (for our case the orbital separation $a$), and $b_{\rm min}$ is defined as the impact parameter for a 90$^\circ$ deflection \cite{2008gady.book.....B}
\begin{equation}\label{eq:bmax}
b_{\rm max}\approx a, \qquad b_{\rm min}={\rm max}(r_h,R_A),
\end{equation}
where $r_h$ is the half-mass radius of the subject system and $R_A$ is the radius at which a particle of the surrounding medium is affected by the sphere of gravitational influence of the test body, namely:
\begin{equation}\label{sec3:a1}
R_{A,i}=\frac{G m_i}{\tilde{v}_{i}^2}.
\end{equation}
For comparison purposes, we shall take values of $b_{\rm max}$ and $b_{\rm min }$ so that $\log{\Lambda}\approx 20$. 

As discussed in Ref.~\cite{2015PhRvD..92l3530P}, we deal with binary systems with orbital velocities smaller than the velocity dispersion of the DM medium so that each binary component does not interact with its respective companion's wake. We are limited then to consider only binary systems with large orbital periods which are, as we shall see, more sensitive to the dynamical friction force, making them more attractive targets to test this effect from observations. Thus, we can apply this approach under the above conditions to binary systems such as NS-NS/NS-WD \cite{2015ASSP...40....1A} and WD-WD \cite{2012ApJ...757L..21H}. Some important assumptions and considerations are summarized to ensure the viability of this approach as follows:
\begin{itemize}
\item the medium is treated as a collisionless gas as long as the mean free path of the DM background is much larger than the size the objects \cite{1990ApJ...359..427B}
\item the Chandrasekhar's DF formula neglects the interaction of one component with its companion’s wake. This condition is guarantee provided that the orbital velocity is smaller that the velocity dispersion of the DM background such that the wake of one object disperses before the arrival of its companion \cite{2015PhRvD..92l3530P}. This also implies that we must consider only binaries with large orbital periods $\dot{P_{b}}>\mathcal{O}(1)$~days
\item linearity of the Chandrasekhar's DF formula requires $L/a \ll 1$, where $L$ is the size of the component’s wake and $a$ is the orbital separation. Here $L$ is determined by Eqn.~(\ref{sec3:a1}) (which is identified as $R_{A}$) and is defined as the radius of the sphere of gravitational influence of the test body
\item the latter condition leads that binary stars can be considered as point masses
\end{itemize}

\section{Orbital decay of binary pulsars by dynamical friction}\label{sec:4}

We review the general approach underlying the DM dynamical friction effect on the orbital period evolution of binary systems. We follow the derivation of the perturbed Keplerian orbits, from the  osculating formalism \cite{2014grav.book.....P} as was presented in Ref.~\cite{2015PhRvD..92l3530P}, to compute the orbital period decay due to the aforementioned effect. We start by writing the equation of motion for the two body system considering the effects of the dynamical friction force of the DM background
\begin{equation}
    m_{i}\ddot{\textbf{r}}_{i}=\pm \frac{G m_{1} m_{2}}{r^{3}}\textbf{r}+\textbf{f}_{fr,i},
\end{equation}
where $\textbf{r}=\textbf{r}_{2}-\textbf{r}_{1}$. Accordingly the relative acceleration between two bodies is written as
\begin{equation}
\dot{\textbf{v}}=-\frac{GM}{r^{3}}\textbf{r}+\textbf{f},\label{c1}
\end{equation}
where $\textbf{v}=\dot{\textbf{r}}$ and $\textbf{f}=a_{1}\eta \textbf{v}+a_{2} \textbf{v}_{w}$ is the perturbing dynamical friction force measured on the center of mass. Note that in the absence of this force, the orbital velocity obeys certainly a Keplerian motion $v=\Omega_{0}r_{0}$, with $\Omega_{0}$ and $r_{0}$ being the angular velocity and orbital separation, respectively. We have also introduced compact definitions: $\eta=\mu/M$, $\mu= m_{p}m_{c}/M$ and $M=m_{p}+m_{c}$, to describe the binary system. Here, the subscripts $p$ and $c$ refer to the primary component and its companion, respectively. With this in mind, the perturbed orbital elements, needed to compute the orbital motion due to the drag force, can be then computed. In particular, the expression for the change of the orbital separation reads
\begin{equation}
\dot{a}=2\sqrt{\frac {r_{0}^{3}}{GM}} S(t),\label{c2}
\end{equation}
%
%
%
%
where $a$ is the semiaxis major. It is important to highlight that the source term $S(t)$,  in the perturbed orbital elements have been defined in terms of the dynamical friction force as well as the wind velocity vector according to Ref.~\cite{2015PhRvD..92l3530P}
\begin{equation}
S(t)=a_{1}\eta v-a_{2} v_{w} \sin\beta \sin(\Omega_{0}t-\alpha),
\end{equation}
%
%
%
In short, the drag force produces a change of the orbital separation with time leading to an intrinsic change of the orbital period $P_{b}=2\pi/\Omega_0$ given by \cite{1980stph.book.....L}
\begin{equation}
\frac{\dot{P_{b}}}{P_{b}}=\frac{3}{2}\frac{\dot{a}}{r_{0}}.\label{c6}
\end{equation}
This simple relation along with the corresponding one for the change of the orbital separation Eqn.~(\ref{c2}) yields the expression for the orbital period time-derivative
\begin{equation}
\dot{P}_{b}(t)=3 P_{b}[a_{1}\eta-a_{2}\Gamma \sin{\beta}\sin{(\Omega_{0}t-\alpha)}].\label{c7}
\end{equation}
Importantly, the measured secular change in the orbital period is obtained after averaging over one period $P_{b}$, namely
\begin{equation}
\langle\dot{P}_{b}\rangle=\frac{1}{P_{b}}\int_{0}^{P_{b}} \dot{P_{b}}(t) dt.
\end{equation}
What concerns to the overall formulation, we have introduced the same definitions as in Ref.~\cite{2015PhRvD..92l3530P} for both an easier comparison of the results and simple implementation of the DM model in consideration $\Gamma=v_{w}/v$, $\Delta_{\pm}=\Delta\pm 1$, $\Delta=\sqrt{1-4\eta}$. The coefficients $a_i$ can be written in terms of the integral velocity contribution function\footnote{The quantity $I_{i}$ is defined by the term in parenthesis in Eqn.~(\ref{a01}).}
\begin{equation}\label{eq:bi}
b_i=\frac{I_{i}}{\tilde{v}_{i}^{3}},
\end{equation}
as
\begin{equation}
a_{1}=-A(b_{1}+b_{2}),\qquad a_{2}=\frac{A}{2}( b_{1}\Delta_{+}+b_{2}\Delta_{-}),\label{c8}
\end{equation}
with $A=4\pi G^{2}M \rho(r)$ and $\rho(r)$ is the mass density of a degenerate gas of free fermions found in Sec.\ref{sec:2}. The dependence of the dynamical friction force on the fermion mass is intrinsically inhered through its relation with the density profile. This incorporation is justified from the fact that we are treating with a gas of self-gravitating fermions which take into account microphysical properties of a system such as the particle mass to describe the mean density. Note that in the above definitions we have considered for simplicity that the Coulomb logarithm in the dynamical friction force is equal for the pulsar and its companion which is a good approximation for $v_{w}>100$~km~ s$^{-1}$. We realize also that different values of the initial phase $\alpha$ does not introduce significant changes in our results, then setting $\alpha=0$ is a suitable choice.

Based on the above formulation, we compute in the next section the secular change of $\dot{P}_b$ due to DM dynamical friction solely. Thus, for the sake of generality of our conclusions, we shall not include additional effects that may contribute to the secular change of the orbital period such as mass loss of the star components or accretion of DM particles onto the binary components. The justification relies on the fact that we are interested in long binaries periods, low-mass and compact star binaries in whose case the formation of accretion disk is unlikely either by Roche-lobe overflow or stellar winds. Hence accretion of matter from one component into the other could occur only via Roche lobe overflow for extremely short binary periods near the merging process. We are also ignoring the effects of DM accretion because of the unknown cross section between DM and baryonic matter inside the stars. For interesting discussion on this latter issue in binary systems see Ref.~\cite{2013ApJ...774...48M}. Furthermore, we are not interested in binaries located in globular clusters where stellar interactions with encounters are more likely neither in regions inside galaxies where gas and dust may perturb the binary system and induce an orbital decay \cite{Prager:2016puh}.

\section{Numerical results: secular change in the orbital period}\label{sec:5}

Let us now compute the orbital period time-derivative according to the underlying description presented in Secs.~(\ref{sec:2})-(\ref{sec:4}). To do so, firstly, we fix two of the three free parameters: $v_{w}$, $P_{b}$ and $m_{f}$ (as it shall be specified in each plot) and then perform the numerical computation varying the remaining one in the pertinent range. This is: $10$~km~s$^{-1}$ $\lesssim v_{w}\lesssim 1000$~km~s$^{-1}$, $0.1$~days $\lesssim 
P_{b} \lesssim 100$~days and $50$~eV $\lesssim m_{f}\lesssim$ 400~eV. The latter range is thus set because it describes properly galactic halos  \cite{2015JCAP...01..002D} and the structural properties of dwarf galaxies  \cite{2015JCAP...01..002D,2017MNRAS.467.1515R,2018MNRAS.475.5385D} being then of astrophysical interest for the purpose of this work. Thus, for $m_{f}=90,200$~eV, their associated central densities shall be $\rho_{0}=0.355,5.6$~GeV cm$^{-3}$. The former value is within the suitable range to fit the Milky-Way Galactic observations \cite{2018arXiv181111125B} while the latter one corresponds roughly to the one inferred to describe a typical dwarf galaxy \cite{2015JCAP...01..002D}. We consider also for the sake of example the following binary system configurations: NS-WD with $m_p=1.3~M_\odot$ and $m_c=0.2~M_\odot$, NS-NS with $m_p=m_c=1.3~M_\odot$ and WD-WD with $m_p=0.5~M_\odot$ and $m_c=0.25~M_\odot$. 
\begin{figure}[h]
\centering
\includegraphics[width=1.0\hsize,clip]{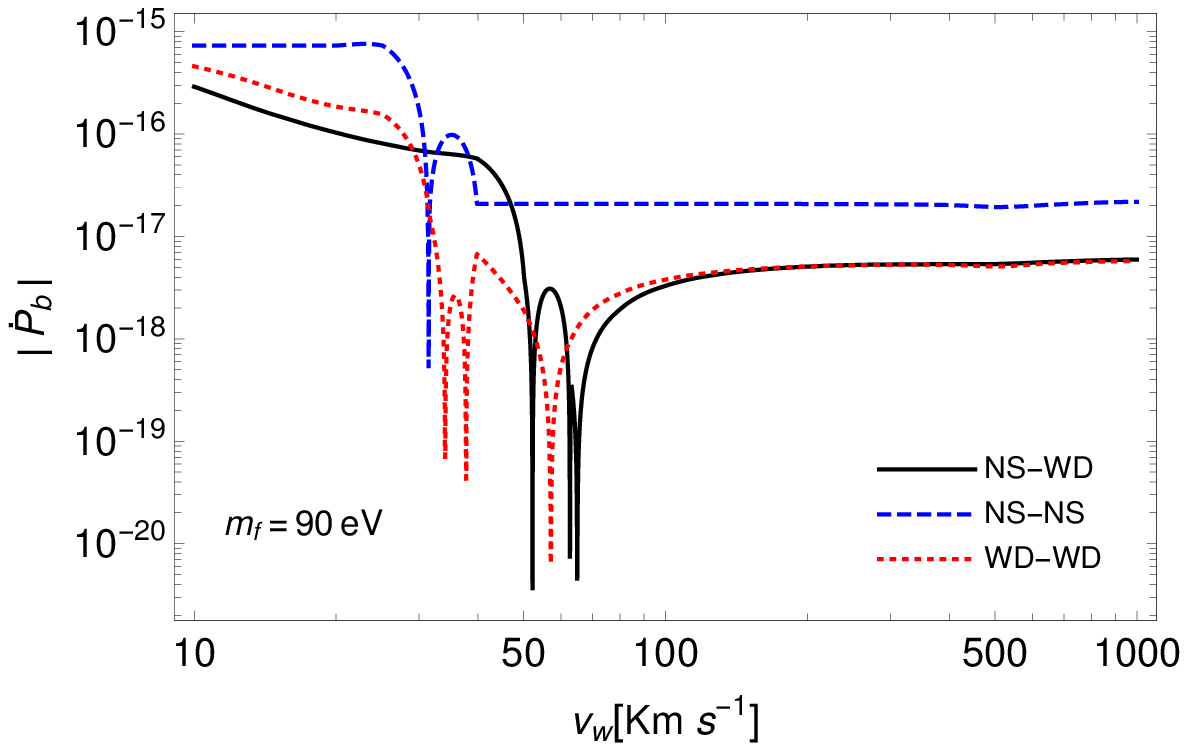}
\includegraphics[width=1.0\hsize,clip]{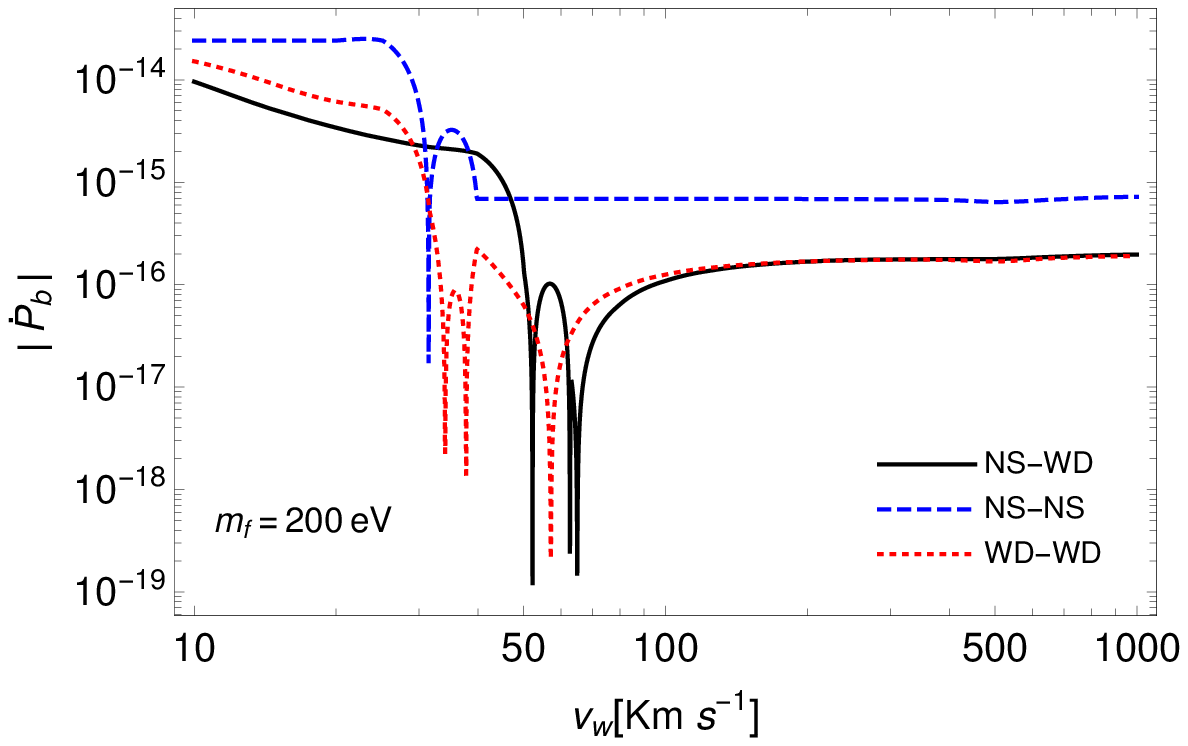}
\caption{Secular change of the orbital period as a function of the DM velocity wind. For this computation we have adopted the values $P_{b}=100$~days and $\beta=\pi/2$. The curves refer to numerical solutions for different binary systems as indicated below: $1.3$--$0.2$ $M_\odot$ NS-WD (black solid line), $1.3$--$1.3$~$M_\odot$ NS-NS (blue dashed line) and $0.25$--$0.50$~$M_\odot$ WD-WD (red dotted line). Top Panel:  This solution corresponds to $m_{f}=90$~eV and $\rho_{0}=0.355$~GeV cm$^{-3}$. Bottom panel: This solution corresponds to $m_{f}=200$~eV and $\rho_{0}=5.6$~GeV cm$^{-3}$.}\label{fig:pbd-vw}
\end{figure}
\subsection{The effect of the wind velocity}

Having established the range of orbital parameters and fermion mass, we start our analysis by investigating the effects of the wind velocity on the secular change of $P_{b}$ for different binary configurations. The results are shown in Fig.~\ref{fig:pbd-vw}. The two plots correspond to a fermion mass of $m_{f}=90$~eV with central density $\rho_{0}=0.355$~GeV cm$^{3}$ (top panel) and $m_{f}=200$~eV with $\rho_{0}=5.6$~GeV cm$^{3}$ (bottom panel), respectively. We can see that $\dot{P}_{b}$ does not exhibit large changes in magnitude (in logarithmic scale) from $v_{w}\approx 40$~~km~s$^{-1}$ for the NS-NS binary and from  $v_{w}\approx 100$~~km~s$^{-1}$ for both  NS-WD and WD-WD binaries. It is also appreciable multiple dips for a given binary star configuration. The physical meaning of those dips in all the plots is due to the change of sign of the orbital period time derivative leading to an orbital shrinking to widening or vice-versa, as appropriate. Such changes of sign occur for certain values of  the binary parameters and the angle $\beta$. As a general trend, we realize that the trivial choice $\beta=0$ does not provide such changes of sign. On the contrary, $\beta=\pi/2$ leads to multiple changes of sign as was also identified in Refs.~\cite{2015PhRvD..92l3530P,2017PhRvD..96f3001G}. We are not interested however in exploring the ranges of the parameters under which such changes occur as was investigated in Ref.~\cite{2017PhRvD..96f3001G} for different DM models. Nevertheless, for the sake of generality of our conclusions, we shall take henceforth $\beta=\pi/2$ in all our numerical estimations.

Taking advantage on this preliminary result, let us do a short commentary beforehand, on the effect of the fermion mass on $\dot{P}_{b}$. We can infer, from Fig.~\ref{fig:pbd-vw}, that it acts only as a scaling factor over all the solutions. Hence, the displayed solutions in both panels of Fig.~\ref{fig:pbd-vw} preserve the same behavior for any binary configuration as can be simply seen. However, for the solution $m_{f}=200$~eV (bottom panel), which reproduces a typical dwarf galaxy according to the mass-radius relation Eqn.~(\ref{eqn6}), the DM dynamical friction effect is almost one order of magnitude larger that the one found for the Milky-Way, i.e. for the solution  $m_{f}=90$~eV. This result yields us some insights about the most promising scenario to search for such an effect in host galaxies.
\begin{figure}[h]
\centering
\includegraphics[width=1.0\hsize,clip]{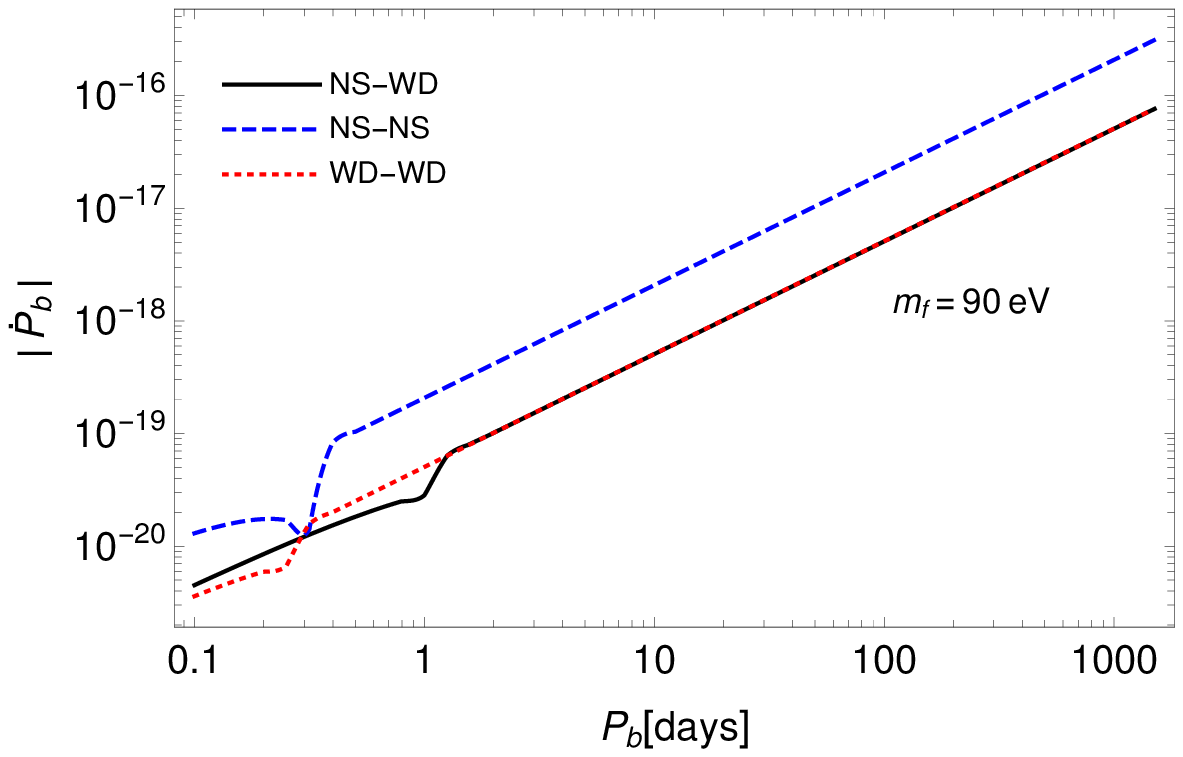}
\includegraphics[width=1.0\hsize,clip]{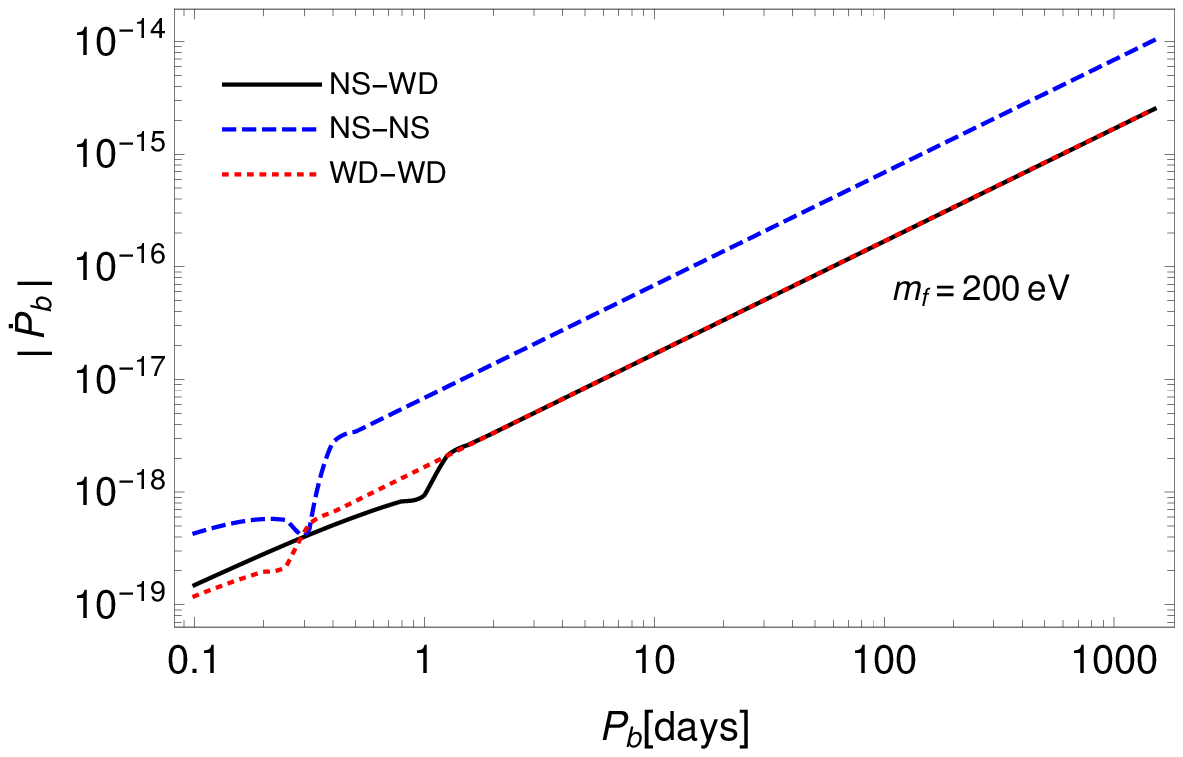}
\caption{Secular change in the orbital period as a function of the orbital period. We have here adopted the values $v_w=100$~km~s$^{-1}$ and $\beta=\pi/2$. The curves refer to numerical solutions for different binary systems as indicated below: $1.3$--$0.2$ $M_\odot$ NS-WD (black solid line), $1.3$--$1.3$~$M_\odot$ NS-NS (blue dashed line) and $0.25$--$0.50$~$M_\odot$ WD-WD (red dotted line). For these solutions we have initially fixed the values of the fermion mass: $m_{f}=90$~eV with $\rho_{0}=0.355$~GeV cm$^{-3}$ (top panel) and $m_{f}=200$~eV with $\rho_{0}=5.6$~GeV cm$^{-3}$ (bottom panel).}\label{fig:pbd-pb}
\end{figure}
\subsection{The effect of the orbital period}

Before proceeding with our analysis, we remind that the two chosen values of fermion masses, adopted here for a self-gravitating degenerate gas of free fermions, lead to a suitable description of both large ($m_{f}\sim 90$~eV) and small ($m_{f}\sim 200$~eV) galaxies. Moreover, we should take in mind these values only as a reference ones since these might change when, for instance, the effect of a thermal envelope is incorporated in the hydrostatic solutions \cite{2017MNRAS.467.1515R} or when additional observational data is included to find the best-fit model \cite{2018arXiv181111125B}. Hence, whichever the fermion mass is, it must describe self-consistently the galaxy size that hosts the binary pulsar.

Turning now to our main discussion, we seek for the dependence of the orbital period time-derivative on the orbital period. We illustrate this in Fig.~\ref{fig:pbd-pb} for the two particular choices of fermion masses $m_{f}=90$~eV (top panel) and $m_{f}=200$~eV (bottom panel) as indicated for different binary star configurations. As expected for the DM dynamical friction effect, the larger the orbital period the larger orbital period time-derivative in agreement with former studies \cite{2015PhRvD..92l3530P,2017PhRvD..96f3001G}. Notice also that NS-NS binary systems experience from $P_{b}\gtrsim0.4$~days larger dynamical friction and therefore larger orbital decay while for NS-WD and WD-WD binaries, their orbital period time-derivatives are essentially equal from $P_{b}\gtrsim 1$~days and smaller than the ones given by NS-NS binary systems for a fixed fermion mass. These features can be observed in Fig.~\ref{fig:pbd-pb}.

Another crucial point, aiming to find an ideal scenario to test the DM dynamical friction effect, is the fact that such an effect could be observed, with major astrophysical expectations, in binaries with large orbital periods which is also a challenge of outstanding precision of pulsar observations. 

\subsection{The effect of the DM fermion mass}

Finally, we compute the secular change of $P_{b}$ with respect to the mass of the degenerate fermion. To do so, we firstly have to set for the fermion masses their associated central densities $\rho_{0}$ whose associated configurations may describe the structure
of galactic halos. As mentioned, once the value of $(m_{f},\rho_{0})$ is set, it defines the size of the DM halo that harbor the binary pulsar.

A first systematic approach could be to establish it from the mass-radius relation Eqn.~(\ref{eqn6}) but we find this is a weak correlation despite it describes successfully the galactic DM sizes. Instead, we use a much reliable strategy as was outlined in \cite{2018MNRAS.475.5385D} by using further astrophysical constraints such as the halo radius and the (nearly constant) central surface DM density of galaxies \cite{2009MNRAS.397.1169D}. This is $\Sigma_{0}=\rho_{0}R(m_{f},\rho_{0})$ with $\Sigma_{0}\approx 100$~$M_{\odot}$ pc$^{-2}$. From this scaling relation, we find the $(m_{f},\rho_{0})$ parameter space consistent with the desire configurations such that we can estimate more reliably the orbital decay of binary stars depending on the type of their host galaxies.

\begin{figure}
\centering
\includegraphics[width=1.0\hsize,clip]{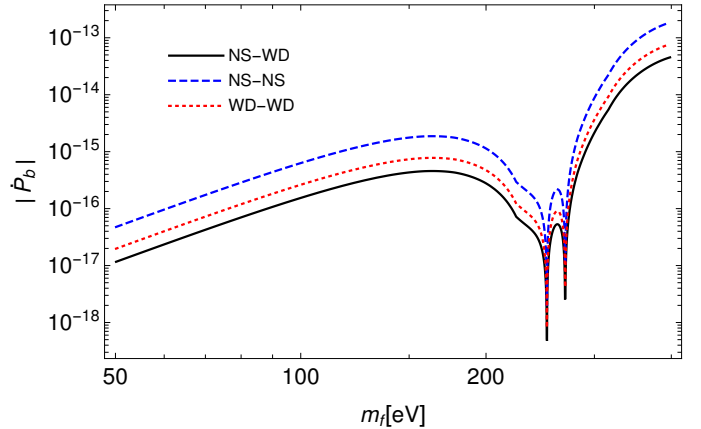}
\caption{Secular change of the orbital period as a function of the fermion mass. As before, we have here adopted the values $v_w=100$~km~s$^{-1}$ and $\beta=\pi/2$. The curves refer to numerical solutions for different binary systems as indicated below: $1.3$--$1.3$ $M_\odot$ NS-NS (top panel), $1.3$--$0.2$~$M_\odot$ NS-WD (middle panel) and $0.25$--$0.50$~$M_\odot$ WD-WD (bottom panel). For this case, we have taken $P_{b}=100$~days. halos.}\label{fig:pbd-mf}
\end{figure}

The result is plotted in Fig.~\ref{fig:pbd-mf} for different binary configurations as before. As we previously advertised, the solutions exhibit a simple scaling dependence for a given star configuration. We can see readily that these solutions are in agreement with the ones displayed in Fig.~\ref{fig:pbd-pb} when $P_{b}=100$~days. Interestingly, NS-NS binaries lead to a slightly  larger (but appreciable) secular changes of $P_{b}$ as can be also appreciated in all the plots. This makes then this binary system an ideal target to test our theoretical predictions. 
For any fixed orbital period, the larger the fermion mass the larger the orbital period time-derivative which implies that this kinematic effect may be larger in binary systems orbiting inside small DM halos as the satellite galaxies of the Milky-Way (e.g. $m_{f}\gtrsim150$~eV). Accordingly, this effect is almost reduced by one order of magnitude in Milky-Way like galaxies (e.g. $m_{f}\sim90$~eV). 
\begin{table*}[]
\begin{tabular}{llllllll}
\multicolumn{1}{c}{Name} & Type  & $m_{p}$ $[M_{\odot}]$ & $m_{c}$ $[M_{\odot}]$ & $P_{b}$[days] & $|\dot{P}_{b}^{obs}|$                      &  $|\dot{P}_{b}|$                      & Ref.                                   \\
\hline
J1740-3052               & NS-NS & 1.4                                     & 2                                       & 231.03             & $3\times10^{-9} $    &  $1.08\times10^{-13}$ & \cite{2012MNRAS.425.2378M} \\
J1713+0747               & NS-WD & 1.31                                    & 0.286                                   & 67.82               & $2.0\times10^{-13}$    & $7.08\times10^{-15}$ & \cite{2015ApJ...809...41Z} \\
J0437-4715               & NS-WD & 1.76                                    & 0.254                                   & 5.741              & $3.73\times10^{-12}$    & $5.67\times10^{-16}$ & \cite{2008ApJ...679..675v}                 \\
J2019+2425               & NS-WD & 1.33                                    & 0.35                                    & 76.51              & $3\times10^{-11}$    & $9.41\times10^{-15}$ & \cite{2001ApJ...549..516N} \\
J1903+0327               & NS-NS & 1.667                                   & 1.029                                   & 95.17              & $53\times10^{-12}$  & $2.77\times10^{-14}$ & \cite{2011MNRAS.412.2763F} \\
\hline
\end{tabular}
\caption{Limits in the measured orbital period time-derivatives for several binary systems in the Galaxy (column 6) \cite{2005AJ....129.1993M} and the ones predicted by DM dynamical friction of degenerate light fermions (column 7). We have set $m_{f}=1$~keV and assumed that the central density $\rho_{0}$ at the binary position can be approximated to the DM density near the sun position, namely $\rho_{0}=0.355$~GeV cm$^{3}$. Here, it is also shown the values of mass binaries (pulsars and their companions) and orbital periods. }\label{tab:pulsars}
\end{table*}
\subsection{Upper limit on the fermion mass from Pulsar timimg observations}

It is instructive at this point to use some available measurements of the orbital period time-derivative for long-period binaries in the Milky-Way to quantify more realistically such an effect, as well as to put constraints on the fermion mass. We evaluate then this effect and show it in Table (\ref{tab:pulsars}) for different binaries-type with measured  orbital period time-derivatives\footnote{These binary stars with long orbital periods have been identified from the ATNF Pulsar Catalogue http://www.atnf.csiro.au/research/pulsar/psrcat/ \cite{2005AJ....129.1993M}.}. For such a computation we have considered a benchmark $m_{f}=1$~keV in accordance with the  bounds derived below. As we have discussed, NS-NS binaries (which correspond to J1740-3052 and J1903+0327 in our case) show slightly larger orbital period time-derivatives due to DM dynamical friction (column 7) and can be used to put stringent bounds on the fermion mass\footnote{Though a direct comparison with the other NS-WD binaries can not be done directly since they differ in orbital periods and orbital period time-derivatives.}. Moreover, it also depends on which configuration of the binary system observed has a comparable orbital period time-derivative with the one predicted by DM dynamical friction. In order to set a bound for the fermion mass from timing pulsar measurements, we demand that the orbital period time derivative due to dynamical friction does not excess the value limit of the observed one for each binary system in consideration. Thus, we look for a maximum value of the fermion mass such that the mentioned condition is satisfied.
For instance, by using J1903+0327 NS-NS measurements we can estimate the upper bound for the fermions mass
\begin{equation}
    m_{f}\lesssim 2~\rm keV,\label{5.1}
\end{equation} 
while the  orbital period time-derivative measured in J1713+0747 NS-WD, which is roughly two order of magnitude smaller than the J1903+0327 binary, yields
\begin{equation}
    m_{f}\lesssim 1~\rm keV,\label{5.2}
\end{equation} 
which are, however, above the ones found in Refs.~\cite{2015JCAP...01..002D,2017MNRAS.467.1515R,2018MNRAS.475.5385D} by means of velocity dispersion measurements. Nevertheless, DM dynamical friction can be considerably increased, as we showed, in binary pulsars orbiting small host galaxies (and/or) with very long orbital periods\footnote{It remains the possibility of considering self-interactions between degenerate DM fermions that may increase the local density and lead to a larger  orbital period time-derivative. This is however beyond the scope of this work.} whereby it would permit us to put more stringent bounds on the fermions mass. It is important however to mention that there has been discovered, up to the best of our
knowledge, around 30 extragalactic pulsars inside satellite galaxies of the Milky-Way, but most of them with unmeasured orbital periods and orbital period time-derivatives\footnote{For more details see Ref.~\cite{2006ApJ...649..235M} and references therein.}.

Despite the current limitations of measuring the orbital period decay of extragalactic pulsars and binary pulsars in the Galaxy with very long orbital periods, there are encouraging prospects for high-precision pulsar timing with the new generation of radio telescopes. In particular, the Square Kilometre Array (SKA) will reach the capability of observing binary pulsars in extreme environments, with improved timing precision, and detecting pulsars in nearby galaxies \cite{2015aska.confE..40K,2018RSPTA.37670293S}. Specifically SKA expects to achieve a precision in the pulse arrival time by a factor of 100 better than current measurements. SKA is also expected to detect very long period pulsars in binary systems and to discover new systems up to $\mathcal{O}(10^3)$ in high DM densities, including binary systems in nearby galaxies within $5-10$~Mpc, depending on the strength of the pulses \cite{Smits:2008cf,2011MNRAS.417.2916L}. Thus, the current bound can be improved up to two orders of magnitude. Hence, we expect to reach the technological improvement in the future that permits us to measure with outstanding precision orbital period decays, under the conditions found, to test our theoretical predictions.

In summary, we have studied in this work the effect of DM dynamical friction in the orbital evolution of binary pulsars in which DM halos are constituted of self-gravitating degenerate gas of free fermions. Thereby, we have accessed to quantify the secular change in the orbital period for this DM model and therefore to devise the best astrophysical scenario to constrain the properties of light fermionic DM by using pulsar timing observations.

\section{Conclusions}

We have assessed the conditions under which DM dynamical friction effect of a degenerate and non-interacting Fermi gas can be potentially tested with the help of new generation of pulsar surveys. The main conclusions of this work are:

\begin{enumerate}

\item DM dynamical friction can be considered as a (an additional) kinematic effect in the observed orbital decay of binary pulsars when high sensitivity of pulsar observations is remarkably attained.

\item We have showed that this effect may become important in future measurements of the orbital period decay in binaries with characteristic long periods, i.e. for $P_{b}\gtrsim 100$~days (see Fig.~\ref{fig:pbd-pb}).

\item  We have quantified the effects of the DM dynamical friction in different binary star configurations. More specifically, we found that NS-NS binary systems experience slightly larger orbital period decays, i.e. $\dot{P}_{b}\gtrsim10^{-15}$ for $P_{b}\gtrsim 100$~days and $m_{f}=200$~eV.

\item We have computed the DM dynamical friction effect for some binary stars in the Milky-Way with measured orbital period time-derivatives (see Table \ref{tab:pulsars}). From the J1713+0747 binary, the constraint on the fermion mass $m_{f}\lesssim 1$~keV have been put. Moreover, this bound can be considerably improved by using timing pulsar measurements of very long-orbital period binary pulsars.

\item We found another promising situation in which binary pulsars orbiting small galaxies, which corresponds to degenerate fermionic DM halos with $m_{f}\sim 200$~eV, experience larger orbital period decay by dynamical friction because of the enhancement of this effect with the fermion mass (see Figs.~\ref{fig:pbd-vw}-\ref{fig:pbd-mf}). Thus, the fermion mass establishes the galaxy size that harbor the binary pulsar. Hence, the established bound in Eqn.~(\ref{5.2}) for the fermion mass can also be improved in such small host galaxies. Interestingly, the SKA's sensitivity will be sufficient to detect pulsars in nearby galaxies which makes of our results testable predictions in the near future \cite{2015aska.confE..40K,2018RSPTA.37670293S}.

\item  It has been shown that there are distinctive theoretical predictions of the orbital period time-derivative for certain DM model as the ones found in this work. Hence, we are reaching a phenomenological situation in which we may constrain or even rule out DM models by using pulsar timing measurements.

\end{enumerate}

DM dynamical friction is an appealing effect on the binary evolution because it would permit us to put constraints on the local DM environment the binaries are embedded due to the high-precision measurements which is a characteristic property in such systems. We found as the main conclusion of this work that NS-NS binary systems with large orbital periods $P_{b}\gtrsim100$~days orbiting small DM halos, composed of degenerate fermions of $m_{f}\sim200$~eV, (which correspond to extragalactic pulsars), are the best astrophysical scenario to test the effect of dynamical friction of light fermionic DM once observational data of timing pulsar is available. Interestingly, there are promising pulsar surveys that can reach an astonishing sensitivity to test the theoretical predictions based on DM models.

\section*{Acknowledgements}
We thank anonymous referee for all the insightful questions and suggestions that help to improve this manuscript. We would like also to offer our gratitude to Alexander Gallego for careful and meticulous reading of this paper. Special thanks go to Laura Becerra, Nicol\'as Bernal and Cl\'ement Stahl for insightful comments and discussions on different subjects of this work. This project was supported by the  Postdoctoral Fellowship Program N$^\circ$ $	
2018000101$ of the \textit{ Vicerrector\'ia de Investigaci\'on y Extensi\'on}-UIS.

\end{document}